\documentclass[journal=jacsat,manuscript=article]{achemso}

\usepackage[version=3]{mhchem} 
\usepackage{gensymb}
\usepackage{lineno}
\usepackage{amsmath}
\usepackage{float}

\author{Jianyu Yang}
\author{Nan Li}
\email{nanli@zju.edu.cn}
\author{Yuyao Pan}
\affiliation[Zhejiang University]
{State Key Laboratory of Extreme Photonics and Instrumentation, College of Optical Science and Engineering, Zhejiang University, Hangzhou 310027, China}
\author{Jing Yang}
\author{Zhiming Chen}
\affiliation[Zhejiang Lab]
{Research Center for Advanced Computational Sensing and Intelligent Processing, Zhejiang Lab, Hangzhou 310000, China}
\author{Han Cai}
\affiliation[Zhejiang University]
{State Key Laboratory of Extreme Photonics and Instrumentation, College of Optical Science and Engineering, Zhejiang University, Hangzhou 310027, China}
\author{Yuliang Wang}
\author{Chuankun Han}
\affiliation[ACIRI]
{Beijing Aerospace Control Instrument Research Institute, Beijing 100854, China
}
\author{Xingfan Chen}
\author{Cheng Liu}
\author{Huizhu Hu}
\email{huhuizhu2000@zju.edu.cn}
\affiliation[Zhejiang University]
{State Key Laboratory of Extreme Photonics and Instrumentation, College of Optical Science and Engineering, Zhejiang University, Hangzhou 310027, China}

\title
  {From photon momentum transfer to acceleration sensing}

\abbreviations{}

\begin{document}

\begin{abstract}
\label{sec:abs}
As a typical application of photon momentum transfer, optical levitation systems are known for their ideal isolation from mechanical dissipation and thermal noise. These characters offer extraordinary potential for acceleration precision sensing and have attracted extensive attention in both fundamental and applied physics. Although considerable improvements of optical levitation accelerometers has been reported, the dynamic testing of the sensing performance remains a crucial challenge before the utilization in practical application scenarios. In this work, we present a dual-beam optical levitation accelerometer and demonstrate the test with dynamic inputs for the first time. An acceleration sensing sensitivity of $0.1\mu g$ and a measurement range of $ 1g$ are achieved. These advancements solidify the potential of optical levitation accelerometer for deployment in practical domains, including navigation, intelligent driving, and industrial automation, building a bridge between the laboratory systems and real-world applications.
\end{abstract}

\section{Introduction}
\label{sec:intro}
Acceleration, a crucial physical quantity, has been frequently measured in the advancement of science and technologies. As the properties of conventional quartz accelerometers are now limited by the material and the manufacturing processes \cite{20450}, there is a rapid increasing demand of new-type accelerometers across a spectrum of applications. Accelerometers based on MEMS \cite{RN33,RN38}, electrostatic levitation \cite{ele_HUST,science_levi}, and cold atom interferometry \cite{atomreview} have been developed to fit different demanding situations. However, before any new-type accelerometer can be put into real-world applications, it is essential to test its properties with dynamic input \cite{dytest_1,dytest_2}. And only through dynamic input testing can we verify the accelerometers' performance under real-world application in a most straightforward and convincing way. \par
Optical levitation systems, pioneered by Ashkin in 1976 \cite{RN1}, have made great progress in the precision sensing applications \cite{RN4,RN5,RN6,RN7}. Thanks for their exceptional isolation from mechanical and thermal noise, the vacuum optical levitation systems are particularly suitable for acceleration sensing \cite{RN15,RN16,refLiTong}, offering theoretical sensitivities lower than $1 ng$ \cite{MIT2010}, surpassing the state-of-the-art accelerometer systems. At present, however, optical levitation accelerometers are still in the experimental stage inside laboratories, and the journey towards their real-world application  is still in its infancy \cite{ref2021}. The main challenge is that most laboratory systems are now designed for static testing environments and can't maintain their precision under dynamic inputs, or even failing to function. Besides, many optical levitation accelerometers are limited by structure and can't achieve a wide range of attitude adjustments \cite{RN15,RN16}, or their measurement range is less than 1 mg, which all restrict their applications in many situations. Therefore, to bridge the gap between laboratory systems and real-world applications, it is essential to design an optical levitation accelerometer with high sensitivity, large measurement range, and large attitude adjustment ability, whose properties can be verified under dynamic inputs. \par
In this paper, we present a dual-beam optical levitation system that levitate a $25\mu m$ diameter silica microsphere, thereby enabling high-sensitivity acceleration sensing.This system is designed to maintain stability across a large range of attitude adjustments of $\pm 90 \degree$ through a close-loop method. We have conducted a thorough evaluation of the system's sensing properties using dynamic input methods for the first time in optical levitation accelerometer field, which has substantiated the system's capability of achieving a sensing sensitivity of $0.1\mu g$ with a measurement range that extends beyond $\pm 1g$. The outstanding sensing performance of our system build a robust bridge between the laboratory systems and a large spectrum of real-world applications, particularly in aerospace \cite{RN25}, navigation systems \cite{RN26}, intelligent driving \cite{RN27}, industrial automation \cite{RN28}, and the proliferation of smart devices \cite{RN29}.\par

\section{Experimental system}
\label{sec:system}
\begin{figure}[htbp]
\centering\includegraphics[width=12cm]{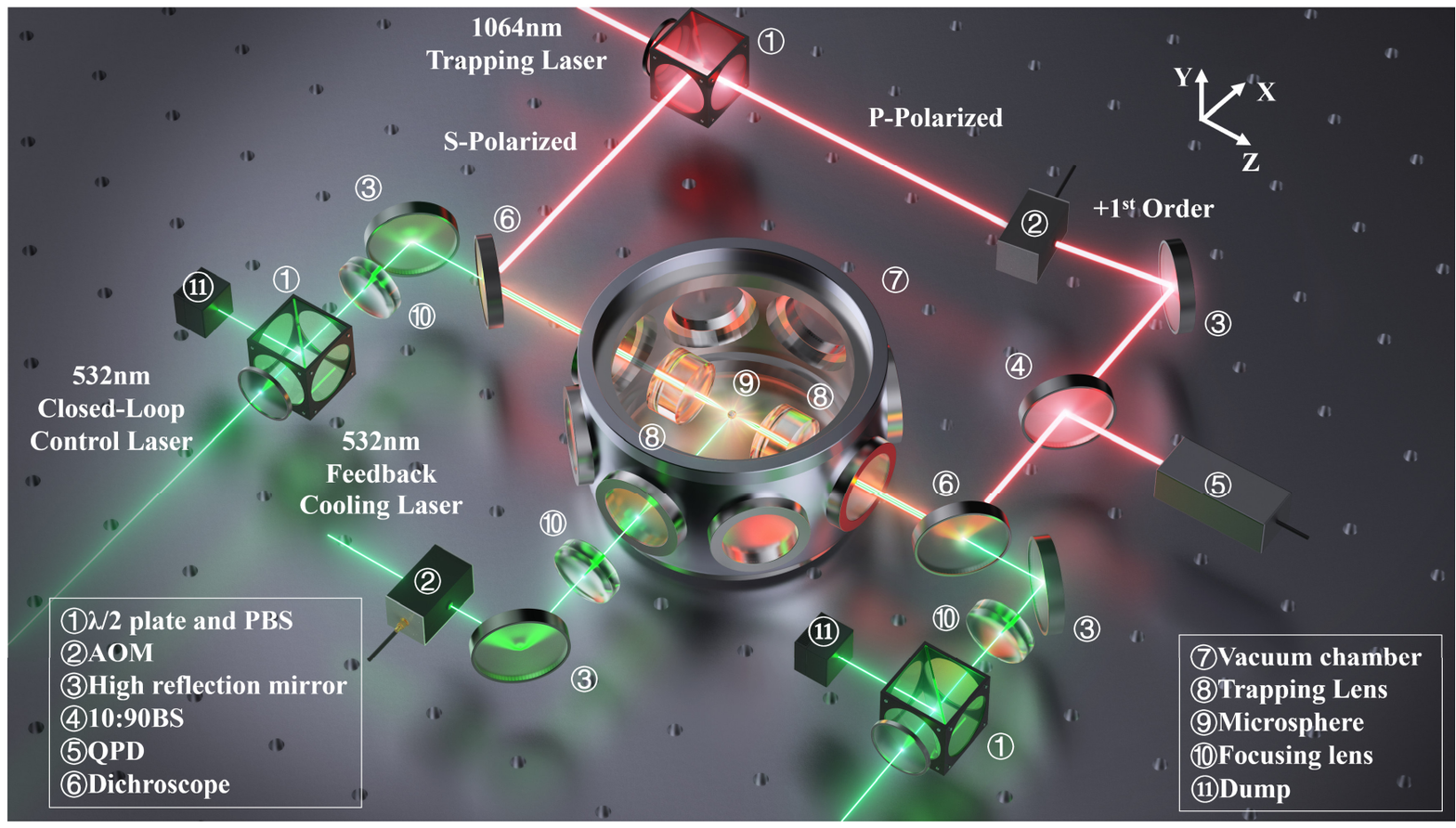}
\caption{Simplified schematic of the experimental setup. A microsphere with a diameter of $25\mu m$ is levitated within a vacuum chamber by a counter-propagating optical trap. Two $532nm$ lasers irradiate the levitated microsphere in the x and y directions, functioning as the feedback cooling beams (y-axis beam has been omitted for a more clearly structrue). Additionally, two high-power $532nm$ laser beams, aligned with the trapping beam, serve as the acceleration modulation beams for measurement range testing.}
\label{fig:system}
\end{figure}
The sensing sensitivity of the optical levitation accelerometer is limited by the following equation \cite{li2012fundamental}:
\begin{equation}
\label{eq:1}
S_{a}=
\sqrt{\frac{4k_{B}T\Gamma }{m}}
\end{equation}\par
Where $k_{B}$ is the Boltzmann constant, T is the microsphere center of mass motion temperature, $\Gamma$ is the damping rate, and m is the mass of the microsphere. The equation suggests that an enhancement in system’s sensitivity can be achieved by increasing the mass of the levitated microsphere. As depicted in Fig. \ref{fig:system}, the experimental setup levitates a silica microsphere with a diameter of $25\mu m$, serving as the sensing oscillator. The microsphere reaches a mass of $18ng$, enabling it to achieve ultra-high acceleration sensing. The system utilizes two $1064nm$ laser beams, focused by two aspheric lenses with an $18.4mm$ focal length within a vacuum chamber, to form a dual-beam optical trap with a Numerical Aperture (NA) of 0.10. This setup is designed to achieve a broader range of attitude adjustments. The $1064nm$ laser beam, after passing through a pointing stabilization system, is then split by a $\lambda/2$ plate and a PBS (Polarizing Beam Splitter) into two beams with orthogonal polarization directions. These beams have a combined power of $400mW$, with a power discrepancy not exceeding 5\% between them. One of the two beams, modified by an AOM (Acousto Optic Modulator), forms the optical trap by the $+1^{st}$ diffracted order. The AOM fulfills two functions: it adjusts the beam’s power for axial (z-direction) feedback cooling and alters the frequency of the beam to prevent interference \cite{RN31}. The s-polarized beam, after passing through the microsphere, is partially reflected by a beam splitter towards a quadrant photodetector, which detects the microsphere’s motion. Two $532nm$ laser beams, modulated in power by AOMs, irradiate the levitated microsphere in the x and y directions, providing feedback cooling for each axis. Moreover, two high-power $532nm$ laser beams are directed towards the levitated microsphere along both axial directions, acting as simulation inputs. The power of these beams is modulated by two electronically controlled $\lambda/2$ plates and PBSs. During experiment, a high-vacuum environment of a pressure of $1 \times 10^{-7} mBar$ will be maintained in the vacuum chamber. According to Equation \ref{eq:1}, the optimal acceleration sensing sensitivity achievable by the system in an ideal noise environment is $5.89ng/\sqrt{\text{Hz}}$.

\section{Measurement range}
\label{sec:range}
\begin{figure}[htbp]
\centering\includegraphics[width=12cm]{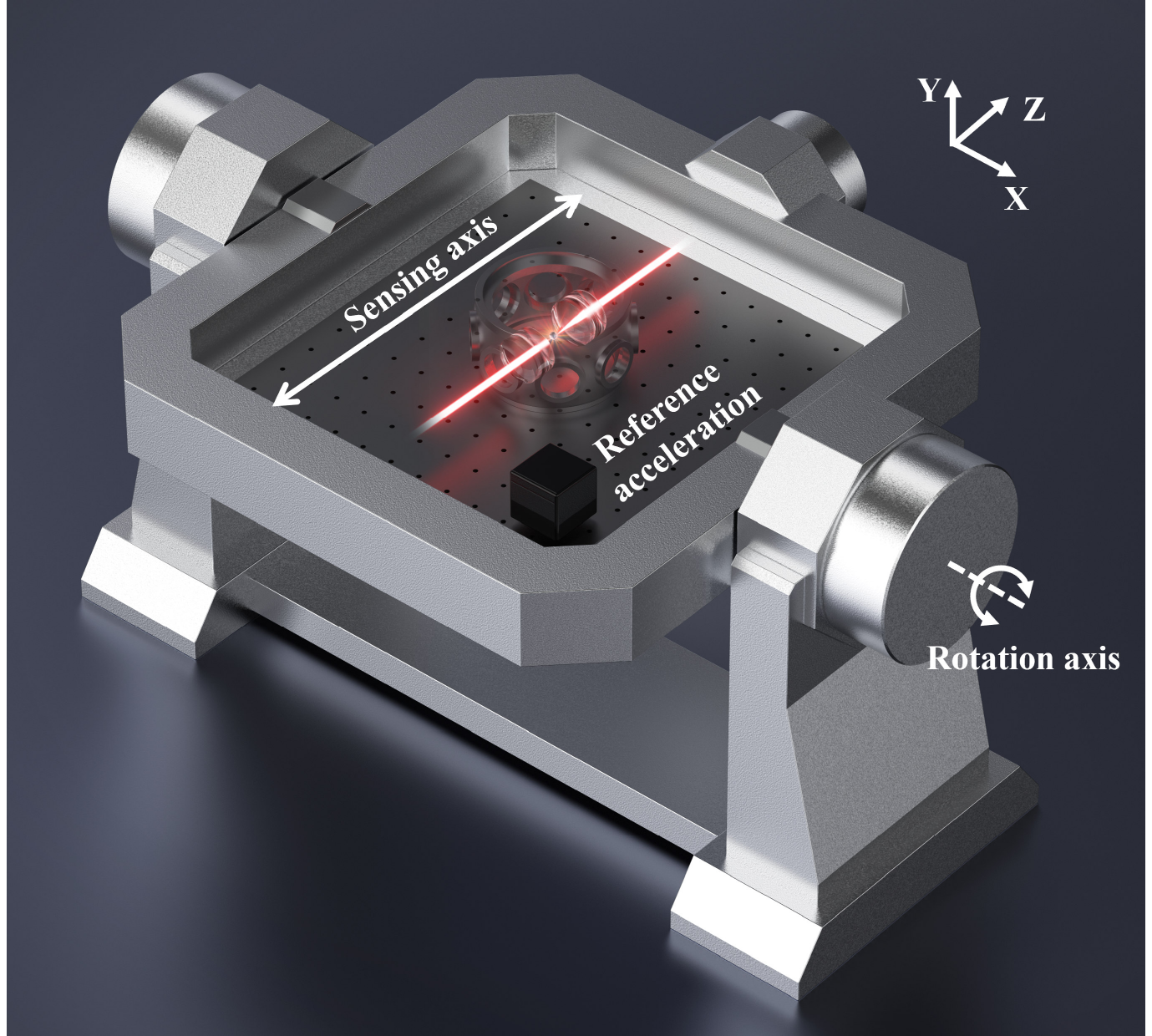}
\caption{The entire optical levitation accelerometer system is mounted on a tilt table, with its rotation axis aligned with the x-axis of the optical system. For reference, an accelerometer is also affixed to the tilt table's edge.}
\label{fig:tilt}
\end{figure}
Upon input of external acceleration to the system, the resulting displacement   of levitated microsphere is governed by the following equation:
\begin{equation}
\label{eq:2}
\Delta x=\frac{4\pi ^{2}a_{\text{in}}}{\Omega ^{2}}
\end{equation}\par
\begin{figure}[htbp]
\centering\includegraphics[width=12cm]{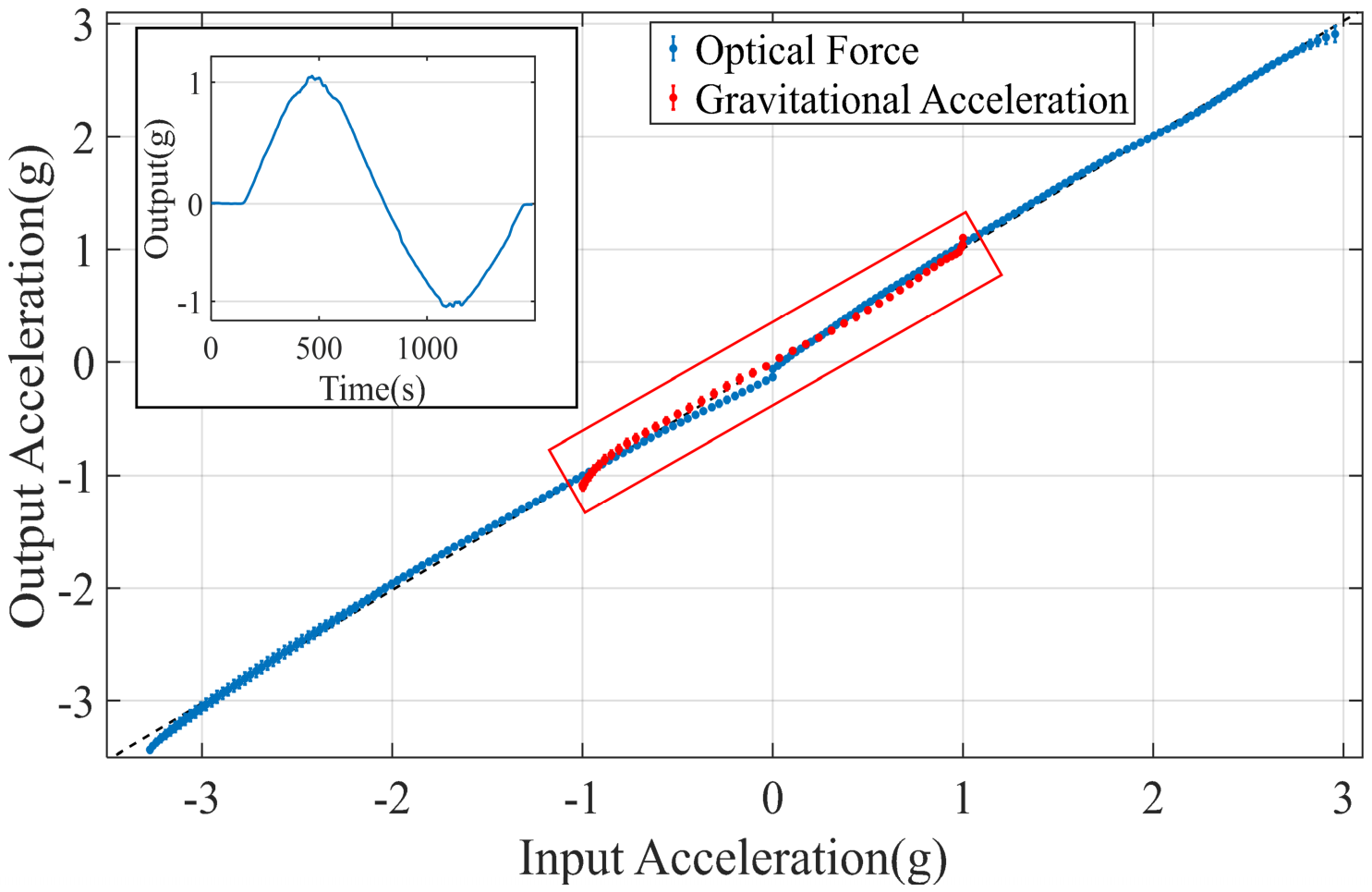}
\caption{Measurement range measurement. The red scatter plot represents the system's input-output curve when the tilt table is rotated to input $\pm 1g$, while the blue scatter plot depicts the input-output relationship curve when a simulated acceleration is applied to the microsphere using high-power $532nm$ laser beams. As the tilt table's rotational input approaches $\pm 1g$, the output error increases due to the rise in optical path pointing noise. This can also be observed in the upper left inset, which shows the system's corresponding signal output curve during the tilt table's rotation. Additionally, the initial state of the high-power $532nm$ laser cannot achieve a 100\% extinction ratio, resulting in a deviation of the simulated input curve near zero.}
\label{fig:range}
\end{figure}

Where $a_{\text{in}}$ is the input acceleration, and $\Omega$ is the natural resonant frequency of the levitated microsphere in the optical trap. According to equation \ref{eq:2}, a system acceleration input of $1g$ ($g$ is the gravitational acceleration) causes the microsphere displacement to exceed $1000\mu m$, significantly surpassing the effective capture range of the optical trap and leading to the microsphere escape. Furthermore, significant deviation of the microsphere from its equilibrium position can lead to obvious nonlinear effects within the detection system. This not only influence the system's sensing performance but also decreases its sensitivity. To address this, we implement a closed-loop control strategy that effectively stabilizes the levitated microsphere close to its equilibrium position. As illustrated in Figure \ref{fig:range}, this control mechanism adjusts the power of the $1064 nm$ trapping beam via an AOM. While the microsphere shifts toward the +z (-z) axis, the AOM correspondingly reduces (increases) the trapping beam's power to recenter the microsphere. Activation of the closed-loop control depends on the microsphere's displacement exceeding ten times the standard deviation of Brownian motion, thus preventing unnecessary interference with the feedback cooling process. With the closed-loop control engaged, the system is capable of maintaining the microsphere's positional fluctuation within ten times the standard deviation of its equilibrium state, even under $\pm 1g$ external acceleration inputs, thereby guaranteeing the system's stability.\par
The introduction of dynamic acceleration input into the system is realized by mounting it on a precision tilt table, as illustrated in Fig. \ref{fig:tilt}. This setup aligns the tilt table’s rotational axis with the system’s x-axis. The alignment is precisely achieved by orienting two apertures, centered on the rotation axis, to coincide with the $532 nm$ feedback cooling beam along x-axis. This alignment, assured by the precision of machining, maintain an error below $0.001\degree $. The tilt table’s rotation, ranging from $-90\degree $ to $+90\degree $, introducing the gravitational acceleration component as the dynamic acceleration input along the axial (z-axis) direction. Fig \ref{fig:range}(red dots) shows the system output variation as the tilt table rotated from the horizontal position ($0g$) to $+90\degree $ ($+1g$), then to $-90\degree$ ($-1g$), and back to $0\degree$. This procedure, performed under a chamber pressure of $1 \times 10^{-7} mBar$ and with rotation parameter set to $0.3\degree /s$ speed and $0.05\degree /s^{2}$ acceleration, ensure the system remains in a quasi-static state to avoid the effects of centripetal acceleration. The system’s output, presenting a sinusoidal respond pattern during rotation, accurately reflects the expected characteristics of the input acceleration, demonstrating the reliability of the closed-loop control mechanism. Additionally, it has also been demonstrated that the system can maintain stability under a wide range of attitude adjustments up to $\pm 90\degree $.\par
To farther ascertain the system’s measurement range, two high-power $532 nm$ laser beam, coaxially aligned with the trapping beam, are introduced along the optical axis to simulate acceleration inputs through optical force on the levitated microsphere. Incrementally increasing the power of one of the green beams until achieving a system output equivalent to $\pm 1g$ facilitated the calibration of the relationship between $532 nm$ laser power and input acceleration. Further increasing the green laser power until the microsphere approaches its escape threshold, after which power reduction and repetition of the procedure on the opposite side afford a evaluation on both sides. The dynamic range of the system, as demonstrated in Fig \ref{fig:range} (blue dots), spans from $-3.2g$ to $+3.5g$, with a linearity of 3\%. The deviation near the zero point is because the high-power $532nm$ laser beam cannot achieve a 100\% extinction through $\lambda/2$ plates and PBSs. The asymmetric dynamic range results from a slight imbalance in the $1064 nm$ trapping beams. \par

\section{Sensitivity}
To evaluate the sensing sensitivity of the system, we input a weak acceleration signal into the system and test its response. We employ the method of rotating the tilt table to input acceleration signals, too. However, constrains imposed by the limited control precision of the tilt table necessitated a different approach. A rigid base was affixed directly beneath the edge of the tilt table, separated by a Piezoelectric Transducer (PZT), which, through its deformation, induces subtle rotational adjustment of the tilt table. By modulating the PZT driving voltage, we can achieve a slight modulation of the tilt table's angle, thereby inputting a weak acceleration into the system. According to the calibration, the relationship between the PZT driving voltage $V$ and the input acceleration $a_{in}$ in the system is $a_{in}=\eta V-0.1673,\eta=0.1957m/V\cdot s^{2}$.\par
\begin{figure}[htbp]
\centering\includegraphics[width=12cm]{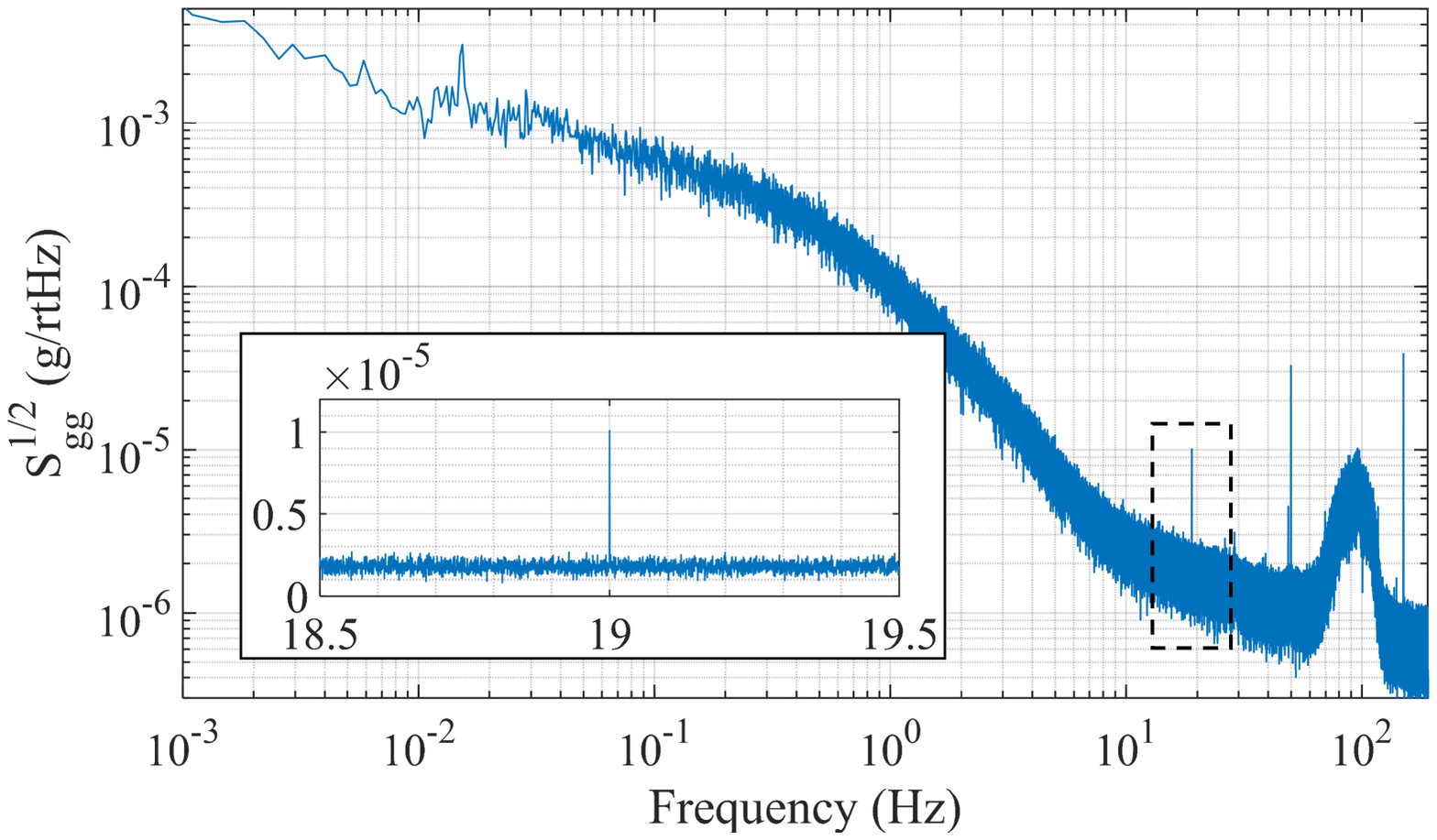}
\caption{Power spectral density of acceleration modulation. By sinusoidally modulating the tilt table, a weak acceleration of $0.1\mu g$ with $19Hz$ in frequency is input into the system, resulting in a distinct motion peak at the modulation frequency.}
\label{fig:PSD}
\end{figure}
Prior to experimentation, a calibration of the detection system was imperative. Leveraging the law of thermal equilibrium, the motion PSD spectrum of the microsphere at a pressure of $10 mBar$, facilitated the calibration of the detection factor to $4.42V/g$ \cite{RN32}. To mitigate the influence of $1/f$ noise and ground vibration noise, the PZT was sinusoidally modulated at $19 Hz$, inputting a continuous sine signal into the system. The alignment of the tilt table axis of rotation with the system x-axis ensure that the microsphere is positioned on the tilt table’s axis, avoiding vertical displacement of the microsphere. Thereby obviate potential interference from extraneous acceleration inputs and guarantee the exclusive input of the gravitational acceleration component. The system’s output was inferred from the peak on the motion PSD spectrum of the microsphere, as delineated in equation \ref{eq:3}:\par
\begin{equation}
\label{eq:3}
a_{\text{output}}=a_{\text{peak}}\sqrt{\frac{2f_{\text{sample}}}{N_\text{FFT}}}
\end{equation}\par
\begin{figure}[htbp]
\centering\includegraphics[width=12cm]{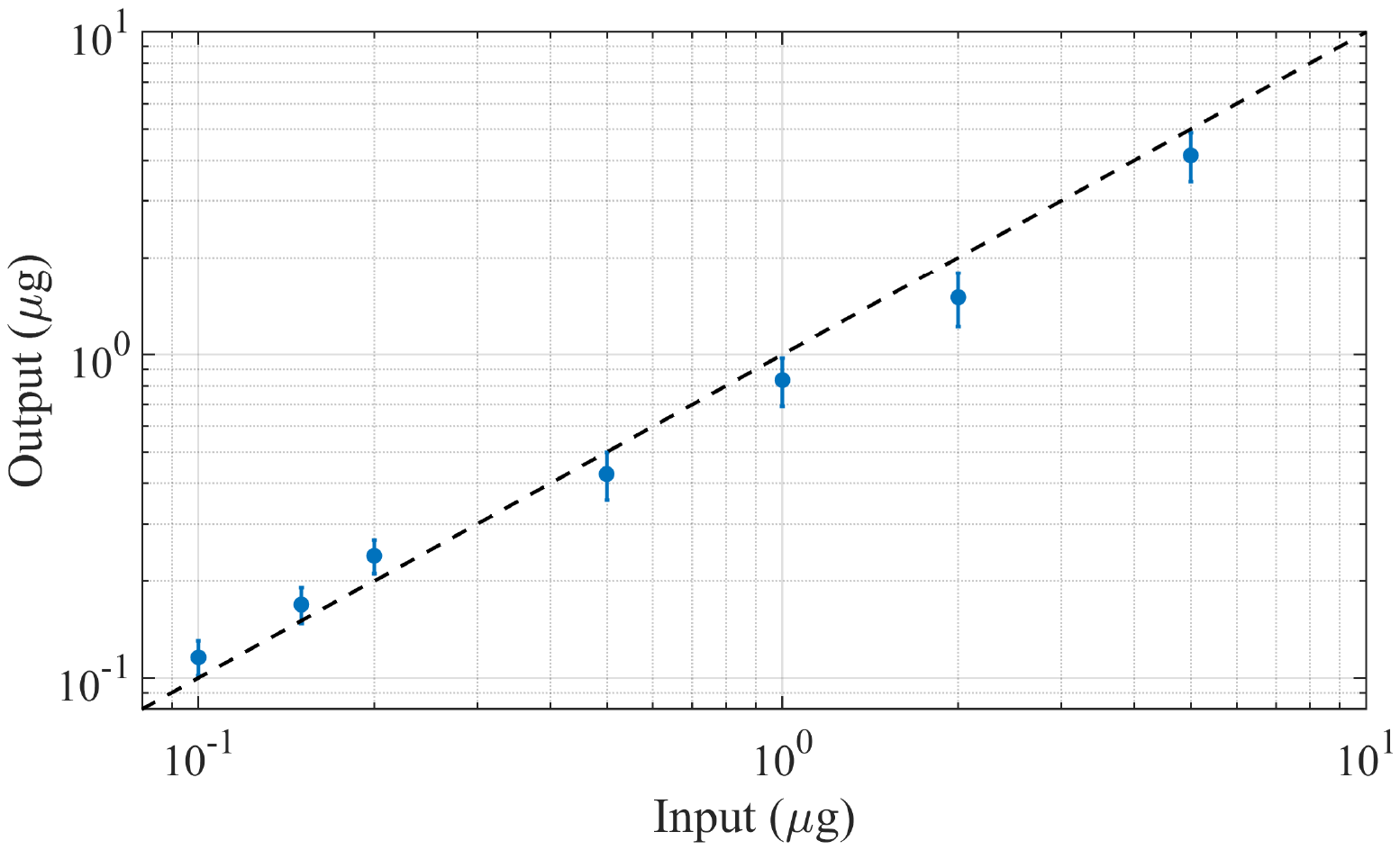}
\caption{System's corresponding output values when inputting different accelerations. The primary source of error in the output is the misalignment between the axis of rotation and the x-axis of the optical trap system.}
\label{fig:inout}
\end{figure}

Where $a_{\text{peak}}$ is the modulated peak intensity of the power spectrum,$f_{\text{sample}}$ is the sampling rate, and ${N_\text{FFT}}$ is the number of the FFT points. As depicted in Fig \ref{fig:PSD}, over a continuous testing period of 10 hours with an input signal of $0.1\mu g$, the motion PSD spectrum of the microsphere was subjected to average filtering at intervals of $2.74\times10^{3}$ seconds, identifying a motion peak at $19Hz$. According to Equation \ref{eq:3}, the output acceleration value can be derived as $0.1157\pm 0.0143(\text{syst.})\pm 0.0037(\text{stat.})\mu g$. Furthermore, as illustrated in Fig \ref{fig:inout}, the linear relationship observed between diminishing input signals and the consequent reduction in system output signals underscores the system sensing reliability. Consequently, the system’s sensitivity in acceleration sensing is $0.1\mu g$.

\section{Discussion and conclusions}
\label{sec:discussion}
\begin{figure}[htbp]
\centering\includegraphics[width=12cm]{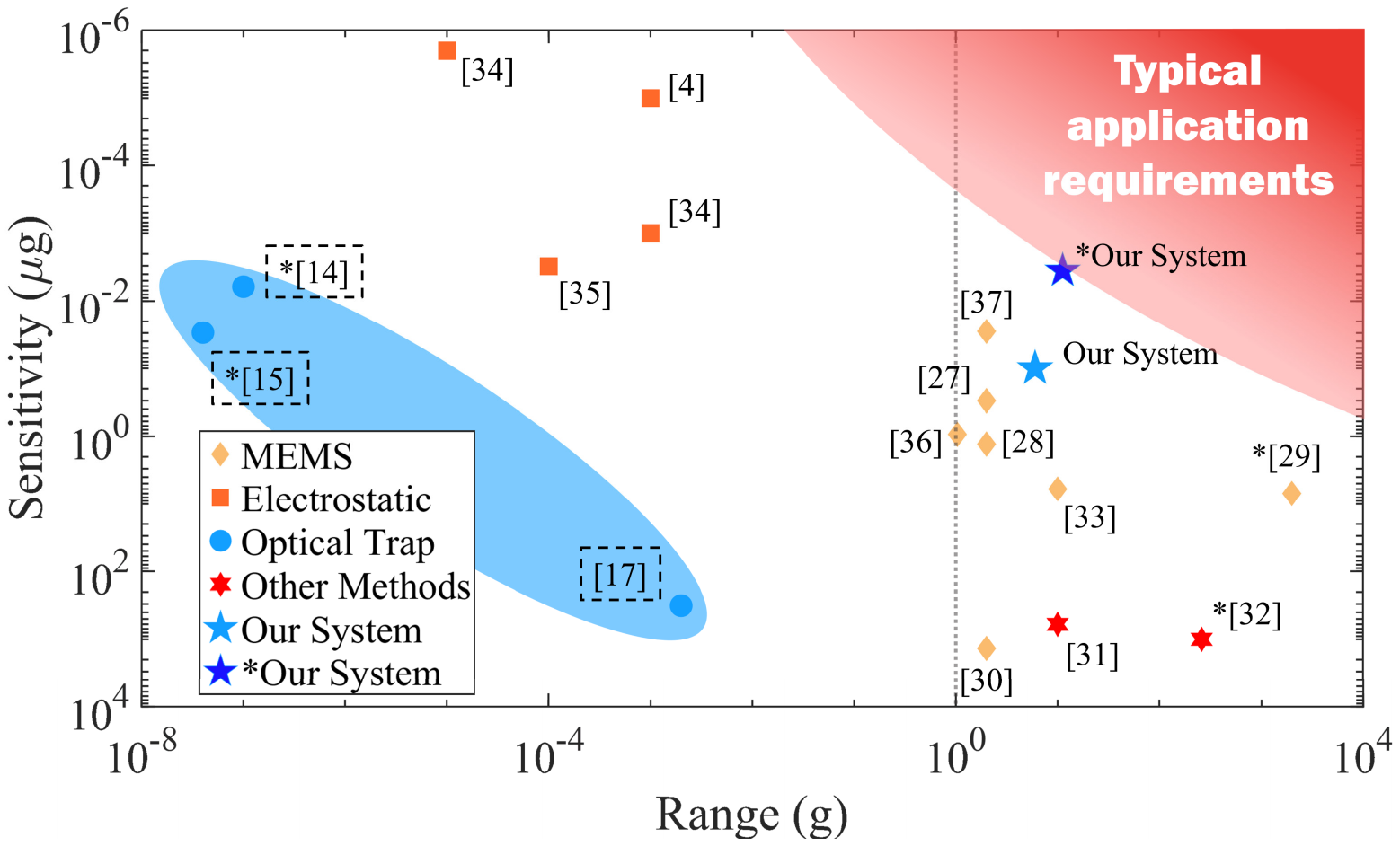}
\caption{Comparison of acceleration measurement range and sensitivity among different systems. The dashed outlines signifying the methods are unable to perform attitude adjustments. And the asterisk (*) in the figure denotes results obtained from simulated calculations. The sensitivity of the optical trap methods were assessed using the same method as presented in this paper.}
\label{fig:compare}
\end{figure}

To actualize the application of optical levitation accelerometers in applications such as navigation, intelligent driving, and industrial automation, it is essential that these systems not only exhibit high sensing sensitivity but also encompass a suitable measurement range and a broad spectrum of attitude adjustments.Figure \ref{fig:compare} delineates a comparative analysis between various systems utilized for acceleration sensing \cite{RN15,RN16,RN33,RN34,RN35,RN36,RN37,RN38,RN39,RN40,RN41,Douch_2014,FLURY20081414,Wang2020,MIT2010,ele_HUST,10195870}, with those encased in dashed outlines signifying the methods are unable to perform attitude adjustments. The sensitivity of the optical trap method was accessed using the same method as presented in this paper as equation \ref{eq:1}. Our system stands out by effectively addressing the limitations of high-sensitivity optical traps, which are traditionally constrained by a narrow measurement range and a lack of attitude adjustability, by achieving a balance between high-sensitivity and measurement capabilities. A comparative assessment against the well-established MEMS technology reveals that our system not only matches the measurement range but also surpasses the majority of MEMS systems in terms of sensing sensitivity. Moreover, it holds the promise of a potential for sensing sensitivity that significantly outstrips current standards. This positions our work as a pivotal advancement for propelling the application of optical levitation accelerometer into a broader spectrum of real-world applications across the aforementioned fields and help to fulfill the escalating demand for heightened sensitivity in those acceleration sensing applications.\par
The sensing properties of our system is currently influenced by two factors. The first pertains to the challenge of maintaining the microsphere’s stability during extensive attitude modifications. This challenge arises from the system’s current design, which predominantly employs a spatial optical path configuration with all optical components vertically anchored to an optical plate. This mounting strategy although standard, lead to an increased vibration amplitude of the components when the system’s attitude is significantly altered. This, in turn, amplifies laser pointing noise, adversely impacting the system stability. A promising solution to this challenge lies in the transition to a fiber-optic system, which not only promises to mitigate these stability issues but also offers the advantages of system miniaturization and enhanced robustness.\par
The second challenge involves mitigating the vibrational noise emanating from the tilt table and the ground. The current system’s acceleration sensing precision, although significantly improved, still falls short of its theoretical potential, as delineated by Equation 1. This discrepancy is largely attributable to environmental vibrational noise, which imposes a fundamental limit on the system’s sensitivity. Our current strategy of employing extended testing periods coupled with average filtering has proven effective in reducing noise interference. However, future enhancements are anticipated through the optimization of the tilt table structure and the integration of an isolation system, aimed at minimizing these vibrational effects further. These anticipated improvements are not only expected to elevate the system’s performance to closer align with theoretical predictions but also to expand its applicability across a broader spectrum of research and industrial applications, solidifying its position as a pivotal tool in the advancement of sensitivity. \par
This paper presents a novel optical levitation accelerometer utilizing a dual-beam trap with a $25um$ diameter microsphere, which is stabilized during substantial acceleration inputs through the implementation of a closed-loop control strategy. Utilizing dynamic inputs, we have convincingly confirmed the system's outstanding sensing performance, with an acceleration sensing sensitivity of $0.1\mu g$, and a measurement range that surpasses $\pm 1g$. Furthermore, the system has demonstrated remarkable stability across a wide spectrum of attitude adjustments, up to $\pm 90\degree $, which is a critical requirement for various technological applications. The contributions of this work are significant, as they establish a robust foundation for the practical applications of optical levitation systems with ultra-high acceleration sensing capabilities into a myriad of fields. These include, but are not limited to, aerospace, navigation, intelligent driving, industrial automation, and smart devices. The system's exceptional sensing performance, combined with its stability and broad applicability, positions it as a promising tool for advancing the state-of-the-art in precision sensing technology.

\bibliography{Reference}

\providecommand{\latin}[1]{#1}
\makeatletter
\providecommand{\doi}
  {\begingroup\let\do\@makeother\dospecials
  \catcode`\{=1 \catcode`\}=2 \doi@aux}
\providecommand{\doi@aux}[1]{\endgroup\texttt{#1}}
\makeatother
\providecommand*\mcitethebibliography{\thebibliography}
\csname @ifundefined\endcsname{endmcitethebibliography}  {\let\endmcitethebibliography\endthebibliography}{}
\begin{mcitethebibliography}{38}
\providecommand*\natexlab[1]{#1}
\providecommand*\mciteSetBstSublistMode[1]{}
\providecommand*\mciteSetBstMaxWidthForm[2]{}
\providecommand*\mciteBstWouldAddEndPuncttrue
  {\def\EndOfBibitem{\unskip.}}
\providecommand*\mciteBstWouldAddEndPunctfalse
  {\let\EndOfBibitem\relax}
\providecommand*\mciteSetBstMidEndSepPunct[3]{}
\providecommand*\mciteSetBstSublistLabelBeginEnd[3]{}
\providecommand*\EndOfBibitem{}
\mciteSetBstSublistMode{f}
\mciteSetBstMaxWidthForm{subitem}{(\alph{mcitesubitemcount})}
\mciteSetBstSublistLabelBeginEnd
  {\mcitemaxwidthsubitemform\space}
  {\relax}
  {\relax}

\bibitem[Filler(1988)]{20450}
Filler,~R. The acceleration sensitivity of quartz crystal oscillators: a review. \emph{IEEE Transactions on Ultrasonics, Ferroelectrics, and Frequency Control} \textbf{1988}, \emph{35}, 297--305\relax
\mciteBstWouldAddEndPuncttrue
\mciteSetBstMidEndSepPunct{\mcitedefaultmidpunct}
{\mcitedefaultendpunct}{\mcitedefaultseppunct}\relax
\EndOfBibitem
\bibitem[Malayappan \latin{et~al.}(2022)Malayappan, Lakshmi, Rao, Ramaswamy, and Pattnaik]{RN33}
Malayappan,~B.; Lakshmi,~U.~P.; Rao,~B. V. V. S. N.~P.; Ramaswamy,~K.; Pattnaik,~P.~K. Sensing Techniques and Interrogation Methods in Optical MEMS Accelerometers: A Review. \emph{IEEE Sensors Journal} \textbf{2022}, \emph{22}, 6232--6246\relax
\mciteBstWouldAddEndPuncttrue
\mciteSetBstMidEndSepPunct{\mcitedefaultmidpunct}
{\mcitedefaultendpunct}{\mcitedefaultseppunct}\relax
\EndOfBibitem
\bibitem[Huang \latin{et~al.}(2021)Huang, Nie, Liu, Liu, Cao, Wang, Cheng, Cui, Gao, and Li]{RN38}
Huang,~K.; Nie,~Y.; Liu,~Y.; Liu,~P.; Cao,~L.; Wang,~Q.; Cheng,~L.; Cui,~J.; Gao,~X.; Li,~J. A Proposal for a High-Sensitivity Optical MEMS Accelerometer With a Double-Mode Modulation System. \emph{Journal of Lightwave Technology} \textbf{2021}, \emph{39}, 303--309\relax
\mciteBstWouldAddEndPuncttrue
\mciteSetBstMidEndSepPunct{\mcitedefaultmidpunct}
{\mcitedefaultendpunct}{\mcitedefaultseppunct}\relax
\EndOfBibitem
\bibitem[Bai \latin{et~al.}(2017)Bai, Li, Hu, Liu, Qu, Tan, Tu, Wu, Yin, Li, and Zhou]{ele_HUST}
Bai,~Y.; Li,~Z.; Hu,~M.; Liu,~L.; Qu,~S.; Tan,~D.; Tu,~H.; Wu,~S.; Yin,~H.; Li,~H.; Zhou,~Z. Research and Development of Electrostatic Accelerometers for Space Science Missions at HUST. \emph{Sensors} \textbf{2017}, \emph{17}\relax
\mciteBstWouldAddEndPuncttrue
\mciteSetBstMidEndSepPunct{\mcitedefaultmidpunct}
{\mcitedefaultendpunct}{\mcitedefaultseppunct}\relax
\EndOfBibitem
\bibitem[Gonzalez-Ballestero \latin{et~al.}(2021)Gonzalez-Ballestero, Aspelmeyer, Novotny, Quidant, and Romero-Isart]{science_levi}
Gonzalez-Ballestero,~C.; Aspelmeyer,~M.; Novotny,~L.; Quidant,~R.; Romero-Isart,~O. Levitodynamics: Levitation and control of microscopic objects in vacuum. \emph{Science} \textbf{2021}, \emph{374}, eabg3027\relax
\mciteBstWouldAddEndPuncttrue
\mciteSetBstMidEndSepPunct{\mcitedefaultmidpunct}
{\mcitedefaultendpunct}{\mcitedefaultseppunct}\relax
\EndOfBibitem
\bibitem[Geiger \latin{et~al.}(2020)Geiger, Landragin, Merlet, and Pereira Dos~Santos]{atomreview}
Geiger,~R.; Landragin,~A.; Merlet,~S.; Pereira Dos~Santos,~F. {High-accuracy inertial measurements with cold-atom sensors}. \emph{AVS Quantum Science} \textbf{2020}, \emph{2}, 024702\relax
\mciteBstWouldAddEndPuncttrue
\mciteSetBstMidEndSepPunct{\mcitedefaultmidpunct}
{\mcitedefaultendpunct}{\mcitedefaultseppunct}\relax
\EndOfBibitem
\bibitem[Kelly \latin{et~al.}(2015)Kelly, Murphy, Watsford, Austin, and Rennie]{dytest_1}
Kelly,~S.~J.; Murphy,~A.~J.; Watsford,~M.~L.; Austin,~D.; Rennie,~M. Reliability and Validity of Sports Accelerometers During Static and Dynamic Testing. \emph{International Journal of Sports Physiology and Performance} \textbf{2015}, \emph{10}, 106 -- 111\relax
\mciteBstWouldAddEndPuncttrue
\mciteSetBstMidEndSepPunct{\mcitedefaultmidpunct}
{\mcitedefaultendpunct}{\mcitedefaultseppunct}\relax
\EndOfBibitem
\bibitem[Minghui \latin{et~al.}(2020)Minghui, Jiang, Bai, Wang, and Wei]{dytest_2}
Minghui,~Z.; Jiang,~K.; Bai,~H.; Wang,~H.; Wei,~X. A MEMS based Fabry–Pérot accelerometer with high resolution. \emph{Microsystem Technologies} \textbf{2020}, \emph{26}\relax
\mciteBstWouldAddEndPuncttrue
\mciteSetBstMidEndSepPunct{\mcitedefaultmidpunct}
{\mcitedefaultendpunct}{\mcitedefaultseppunct}\relax
\EndOfBibitem
\bibitem[Ashkin and Dziedzic(1976)Ashkin, and Dziedzic]{RN1}
Ashkin,~A.; Dziedzic,~J.~M. Optical Levitation in High-Vacuum. \emph{Applied Physics Letters} \textbf{1976}, \emph{28}, 333--335\relax
\mciteBstWouldAddEndPuncttrue
\mciteSetBstMidEndSepPunct{\mcitedefaultmidpunct}
{\mcitedefaultendpunct}{\mcitedefaultseppunct}\relax
\EndOfBibitem
\bibitem[Kuang \latin{et~al.}(2020)Kuang, Xiao, Yu, Fan, and Luo]{RN4}
Kuang,~T.~F.; Xiao,~G.~Z.; Yu,~X.~D.; Fan,~Z.~F.; Luo,~H. Characteristic of intracavity optical tweezers in acceleration detection. \emph{2020 7th Ieee International Symposium on Inertial Sensors and Systems (Inertial 2020)} \textbf{2020}, \relax
\mciteBstWouldAddEndPunctfalse
\mciteSetBstMidEndSepPunct{\mcitedefaultmidpunct}
{}{\mcitedefaultseppunct}\relax
\EndOfBibitem
\bibitem[Mamin and Rugar(2001)Mamin, and Rugar]{RN5}
Mamin,~H.~J.; Rugar,~D. Sub-attonewton force detection at millikelvin temperatures. \emph{Applied Physics Letters} \textbf{2001}, \emph{79}, 3358--3360\relax
\mciteBstWouldAddEndPuncttrue
\mciteSetBstMidEndSepPunct{\mcitedefaultmidpunct}
{\mcitedefaultendpunct}{\mcitedefaultseppunct}\relax
\EndOfBibitem
\bibitem[Ranjit \latin{et~al.}(2015)Ranjit, Atherton, Stutz, Cunningham, and Geraci]{RN6}
Ranjit,~G.; Atherton,~D.~P.; Stutz,~J.~H.; Cunningham,~M.; Geraci,~A.~A. Attonewton force detection using microspheres in a dual-beam optical trap in high vacuum. \emph{Physical Review A} \textbf{2015}, \emph{91}\relax
\mciteBstWouldAddEndPuncttrue
\mciteSetBstMidEndSepPunct{\mcitedefaultmidpunct}
{\mcitedefaultendpunct}{\mcitedefaultseppunct}\relax
\EndOfBibitem
\bibitem[Ranjit \latin{et~al.}(2016)Ranjit, Cunningham, Casey, and Geraci]{RN7}
Ranjit,~G.; Cunningham,~M.; Casey,~K.; Geraci,~A.~A. Zeptonewton force sensing with nanospheres in an optical lattice. \emph{Physical Review A} \textbf{2016}, \emph{93}\relax
\mciteBstWouldAddEndPuncttrue
\mciteSetBstMidEndSepPunct{\mcitedefaultmidpunct}
{\mcitedefaultendpunct}{\mcitedefaultseppunct}\relax
\EndOfBibitem
\bibitem[Monteiro \latin{et~al.}(2020)Monteiro, Li, Afek, Li, Mossman, and Moore]{RN15}
Monteiro,~F.; Li,~W.; Afek,~G.; Li,~C.-l.; Mossman,~M.; Moore,~D.~C. Force and acceleration sensing with optically levitated nanogram masses at microkelvin temperatures. \emph{Physical Review A} \textbf{2020}, \emph{101}\relax
\mciteBstWouldAddEndPuncttrue
\mciteSetBstMidEndSepPunct{\mcitedefaultmidpunct}
{\mcitedefaultendpunct}{\mcitedefaultseppunct}\relax
\EndOfBibitem
\bibitem[Monteiro \latin{et~al.}(2017)Monteiro, Ghosh, Fine, and Moore]{RN16}
Monteiro,~F.; Ghosh,~S.; Fine,~A.~G.; Moore,~D.~C. Optical levitation of 10-ng spheres with nano- g acceleration sensitivity. \emph{Physical Review A} \textbf{2017}, \emph{96}\relax
\mciteBstWouldAddEndPuncttrue
\mciteSetBstMidEndSepPunct{\mcitedefaultmidpunct}
{\mcitedefaultendpunct}{\mcitedefaultseppunct}\relax
\EndOfBibitem
\bibitem[Yin \latin{et~al.}(2013)Yin, Geraci, and Li]{refLiTong}
Yin,~Z.-Q.; Geraci,~A.~A.; Li,~T. Optomechanics of Levitated Dielectric Particles. \emph{International Journal of Modern Physics B} \textbf{2013}, \emph{27}\relax
\mciteBstWouldAddEndPuncttrue
\mciteSetBstMidEndSepPunct{\mcitedefaultmidpunct}
{\mcitedefaultendpunct}{\mcitedefaultseppunct}\relax
\EndOfBibitem
\bibitem[Krish(2010)]{MIT2010}
Krish,~K. Toward a demonstration of a light force accelerometer. Ph.D.\ thesis, Massachusetts Institute of Technology, 2010\relax
\mciteBstWouldAddEndPuncttrue
\mciteSetBstMidEndSepPunct{\mcitedefaultmidpunct}
{\mcitedefaultendpunct}{\mcitedefaultseppunct}\relax
\EndOfBibitem
\bibitem[Gieseler \latin{et~al.}(2021)Gieseler, Gomez-Solano, Magazzù, Pérez~Castillo, Pérez~García, Gironella-Torrent, Viader-Godoy, Ritort, Pesce, Arzola, Volke-Sepúlveda, and Volpe]{ref2021}
Gieseler,~J.; Gomez-Solano,~J.~R.; Magazzù,~A.; Pérez~Castillo,~I.; Pérez~García,~L.; Gironella-Torrent,~M.; Viader-Godoy,~X.; Ritort,~F.; Pesce,~G.; Arzola,~A.~V.; Volke-Sepúlveda,~K.; Volpe,~G. Optical tweezers — from calibration to applications: a tutorial. \emph{Advances in Optics and Photonics} \textbf{2021}, \emph{13}\relax
\mciteBstWouldAddEndPuncttrue
\mciteSetBstMidEndSepPunct{\mcitedefaultmidpunct}
{\mcitedefaultendpunct}{\mcitedefaultseppunct}\relax
\EndOfBibitem
\bibitem[Nima~Mahmoodi and Ahmadian(2010)Nima~Mahmoodi, and Ahmadian]{RN25}
Nima~Mahmoodi,~S.; Ahmadian,~M. Modified acceleration feedback for active vibration control of aerospace structures. \emph{Smart Materials and Structures} \textbf{2010}, \emph{19}, 065015\relax
\mciteBstWouldAddEndPuncttrue
\mciteSetBstMidEndSepPunct{\mcitedefaultmidpunct}
{\mcitedefaultendpunct}{\mcitedefaultseppunct}\relax
\EndOfBibitem
\bibitem[Chin-Woo and Sungsu(2005)Chin-Woo, and Sungsu]{RN26}
Chin-Woo,~T.; Sungsu,~P. Design of accelerometer-based inertial navigation systems. \emph{IEEE Transactions on Instrumentation and Measurement} \textbf{2005}, \emph{54}, 2520--2530\relax
\mciteBstWouldAddEndPuncttrue
\mciteSetBstMidEndSepPunct{\mcitedefaultmidpunct}
{\mcitedefaultendpunct}{\mcitedefaultseppunct}\relax
\EndOfBibitem
\bibitem[Li \latin{et~al.}(2019)Li, Zhang, Chen, Dong, and Zhang]{RN27}
Li,~Z.; Zhang,~K.; Chen,~B.; Dong,~Y.; Zhang,~L. Driver identification in intelligent vehicle systems using machine learning algorithms. \emph{IET Intelligent Transport Systems} \textbf{2019}, \emph{13}, 40--47\relax
\mciteBstWouldAddEndPuncttrue
\mciteSetBstMidEndSepPunct{\mcitedefaultmidpunct}
{\mcitedefaultendpunct}{\mcitedefaultseppunct}\relax
\EndOfBibitem
\bibitem[Neto \latin{et~al.}(2019)Neto, Pires, and Moreira]{RN28}
Neto,~P.; Pires,~J.~N.; Moreira,~A.~P. Accelerometer-based control of an industrial robotic arm. RO-MAN 2009 - The 18th IEEE International Symposium on Robot and Human Interactive Communication. 2019; pp 1192--1197\relax
\mciteBstWouldAddEndPuncttrue
\mciteSetBstMidEndSepPunct{\mcitedefaultmidpunct}
{\mcitedefaultendpunct}{\mcitedefaultseppunct}\relax
\EndOfBibitem
\bibitem[Jain and Kanhangad(2018)Jain, and Kanhangad]{RN29}
Jain,~A.; Kanhangad,~V. Human Activity Classification in Smartphones Using Accelerometer and Gyroscope Sensors. \emph{IEEE Sensors Journal} \textbf{2018}, \emph{18}, 1169--1177\relax
\mciteBstWouldAddEndPuncttrue
\mciteSetBstMidEndSepPunct{\mcitedefaultmidpunct}
{\mcitedefaultendpunct}{\mcitedefaultseppunct}\relax
\EndOfBibitem
\bibitem[Li(2012)]{li2012fundamental}
Li,~T. \emph{Fundamental tests of physics with optically trapped microspheres}; Springer Science \& Business Media, 2012\relax
\mciteBstWouldAddEndPuncttrue
\mciteSetBstMidEndSepPunct{\mcitedefaultmidpunct}
{\mcitedefaultendpunct}{\mcitedefaultseppunct}\relax
\EndOfBibitem
\bibitem[Rafferty and Preston(2021)Rafferty, and Preston]{RN31}
Rafferty,~A.; Preston,~T.~C. Trapping positions in a dual-beam optical trap. \emph{Journal of Applied Physics} \textbf{2021}, \emph{130}\relax
\mciteBstWouldAddEndPuncttrue
\mciteSetBstMidEndSepPunct{\mcitedefaultmidpunct}
{\mcitedefaultendpunct}{\mcitedefaultseppunct}\relax
\EndOfBibitem
\bibitem[Hebestreit \latin{et~al.}(2018)Hebestreit, Frimmer, Reimann, Dellago, Ricci, and Novotny]{RN32}
Hebestreit,~E.; Frimmer,~M.; Reimann,~R.; Dellago,~C.; Ricci,~F.; Novotny,~L. Calibration and energy measurement of optically levitated nanoparticle sensors. \emph{Rev Sci Instrum} \textbf{2018}, \emph{89}, 033111\relax
\mciteBstWouldAddEndPuncttrue
\mciteSetBstMidEndSepPunct{\mcitedefaultmidpunct}
{\mcitedefaultendpunct}{\mcitedefaultseppunct}\relax
\EndOfBibitem
\bibitem[Zhao \latin{et~al.}(2020)Zhao, Jiang, Bai, Wang, and Wei]{RN34}
Zhao,~M.; Jiang,~K.; Bai,~H.; Wang,~H.; Wei,~X. A MEMS based Fabry–Pérot accelerometer with high resolution. \emph{Microsystem Technologies} \textbf{2020}, \emph{26}, 1961--1969\relax
\mciteBstWouldAddEndPuncttrue
\mciteSetBstMidEndSepPunct{\mcitedefaultmidpunct}
{\mcitedefaultendpunct}{\mcitedefaultseppunct}\relax
\EndOfBibitem
\bibitem[Gao \latin{et~al.}(2021)Gao, Zhou, Bi, and Feng]{RN35}
Gao,~S.; Zhou,~Z.; Bi,~X.; Feng,~L. A Low Cross-Axis Sensitivity Micro-Grating Accelerometer With Double-Layer Cantilever Beams. \emph{IEEE Sensors Journal} \textbf{2021}, \emph{21}, 16503--16509\relax
\mciteBstWouldAddEndPuncttrue
\mciteSetBstMidEndSepPunct{\mcitedefaultmidpunct}
{\mcitedefaultendpunct}{\mcitedefaultseppunct}\relax
\EndOfBibitem
\bibitem[V. \latin{et~al.}(2019)V., W.~J., B., X., and R.]{RN36}
V.,~R.; W.~J.,~W.; B.,~F.; X.,~R.; R.,~J. Simulation and Design of an Optical Accelerometer. \emph{2019 20th International Conference on Thermal, Mechanical and Multi-Physics Simulation and Experiments in Microelectronics and Microsystems (EuroSimE)} \textbf{2019}, \emph{pp. 1-6}\relax
\mciteBstWouldAddEndPuncttrue
\mciteSetBstMidEndSepPunct{\mcitedefaultmidpunct}
{\mcitedefaultendpunct}{\mcitedefaultseppunct}\relax
\EndOfBibitem
\bibitem[Wu \latin{et~al.}(2015)Wu, Wang, Li, Huang, Ge, and Yu]{RN37}
Wu,~X.; Wang,~X.; Li,~S.; Huang,~S.; Ge,~Q.; Yu,~B. Cantilever Fiber-Optic Accelerometer Based on Modal Interferometer. \emph{IEEE Photonics Technology Letters} \textbf{2015}, \emph{27}, 1632--1635\relax
\mciteBstWouldAddEndPuncttrue
\mciteSetBstMidEndSepPunct{\mcitedefaultmidpunct}
{\mcitedefaultendpunct}{\mcitedefaultseppunct}\relax
\EndOfBibitem
\bibitem[Chau \latin{et~al.}(1996)Chau, Lewis, Zhao, Howe, Bart, and Marcheselli]{RN39}
Chau,~K. H.~L.; Lewis,~S.~R.; Zhao,~Y.; Howe,~R.~T.; Bart,~S.~F.; Marcheselli,~R.~G. An integrated force-balanced capacitive accelerometer for low-g applications. \emph{Sensors and Actuators A: Physical} \textbf{1996}, \emph{54}, 472--476\relax
\mciteBstWouldAddEndPuncttrue
\mciteSetBstMidEndSepPunct{\mcitedefaultmidpunct}
{\mcitedefaultendpunct}{\mcitedefaultseppunct}\relax
\EndOfBibitem
\bibitem[Jian \latin{et~al.}(2017)Jian, Wei, Guo, Hu, Tang, Liu, Zhang, and Sang]{RN40}
Jian,~A.; Wei,~C.; Guo,~L.; Hu,~J.; Tang,~J.; Liu,~J.; Zhang,~X.; Sang,~S. Theoretical Analysis of an Optical Accelerometer Based on Resonant Optical Tunneling Effect. \emph{Sensors (Basel)} \textbf{2017}, \emph{17}\relax
\mciteBstWouldAddEndPuncttrue
\mciteSetBstMidEndSepPunct{\mcitedefaultmidpunct}
{\mcitedefaultendpunct}{\mcitedefaultseppunct}\relax
\EndOfBibitem
\bibitem[Kolli \latin{et~al.}(2021)Kolli, Dudla, and Talabattula]{RN41}
Kolli,~V.~R.; Dudla,~P.; Talabattula,~S. Integrated optical MEMS serially coupled double racetrack resonator based accelerometer. \emph{Optik} \textbf{2021}, \emph{236}, 166583\relax
\mciteBstWouldAddEndPuncttrue
\mciteSetBstMidEndSepPunct{\mcitedefaultmidpunct}
{\mcitedefaultendpunct}{\mcitedefaultseppunct}\relax
\EndOfBibitem
\bibitem[Douch \latin{et~al.}(2014)Douch, Christophe, Foulon, Panet, Pajot-Métivier, and Diament]{Douch_2014}
Douch,~K.; Christophe,~B.; Foulon,~B.; Panet,~I.; Pajot-Métivier,~G.; Diament,~M. Ultra-sensitive electrostatic planar acceleration gradiometer for airborne geophysical surveys. \emph{Measurement Science and Technology} \textbf{2014}, \emph{25}, 105902\relax
\mciteBstWouldAddEndPuncttrue
\mciteSetBstMidEndSepPunct{\mcitedefaultmidpunct}
{\mcitedefaultendpunct}{\mcitedefaultseppunct}\relax
\EndOfBibitem
\bibitem[Flury \latin{et~al.}(2008)Flury, Bettadpur, and Tapley]{FLURY20081414}
Flury,~J.; Bettadpur,~S.; Tapley,~B.~D. Precise accelerometry onboard the GRACE gravity field satellite mission. \emph{Advances in Space Research} \textbf{2008}, \emph{42}, 1414--1423\relax
\mciteBstWouldAddEndPuncttrue
\mciteSetBstMidEndSepPunct{\mcitedefaultmidpunct}
{\mcitedefaultendpunct}{\mcitedefaultseppunct}\relax
\EndOfBibitem
\bibitem[Wang \latin{et~al.}(2020)Wang, Qi, Wang, Li, Li, and Wang]{Wang2020}
Wang,~Y.; Qi,~K.; Wang,~S.; Li,~W.; Li,~Z.; Wang,~Z. Capacitive Sensing and Electrostatic Control System Design and Analysis With a Torsion Pendulum. \emph{IEEE Access} \textbf{2020}, \emph{8}, 1021--1030\relax
\mciteBstWouldAddEndPuncttrue
\mciteSetBstMidEndSepPunct{\mcitedefaultmidpunct}
{\mcitedefaultendpunct}{\mcitedefaultseppunct}\relax
\EndOfBibitem
\bibitem[Qu \latin{et~al.}(2024)Qu, Ouyang, Xiong, Xu, Wang, and Liu]{10195870}
Qu,~Z.; Ouyang,~H.; Xiong,~W.; Xu,~Q.; Wang,~Y.; Liu,~H. A Nano-g MOEMS Accelerometer Featuring Electromagnetic Force Balance With 157-dB Dynamic Range. \emph{IEEE Transactions on Industrial Electronics} \textbf{2024}, \emph{71}, 6418--6426\relax
\mciteBstWouldAddEndPuncttrue
\mciteSetBstMidEndSepPunct{\mcitedefaultmidpunct}
{\mcitedefaultendpunct}{\mcitedefaultseppunct}\relax
\EndOfBibitem
\end{mcitethebibliography}

\end{document}